# The Provenance Problem: LLMs and the Breakdown of Citation Norms

Brian D. Earp,[1] Haotian Yuan,[2] Julian Koplin[3] & Sebastian Porsdam Mann[1,4]

1. Centre for Biomedical Ethics, Yong Loon Li School of Medicine, National University of Singapore, Singapore
2. Georgetown Preparatory School, Bethesda, Maryland, United States
3. Monash Bioethics Centre, Monash University, Melbourne, Victoria, Australia
4. Centre for Advanced Studies in Bioscience Innovation Law (CeBIL), Faculty of Law, University of Copenhagen

**Preface**: The increasing use of generative AI in scientific writing raises urgent questions about attribution and intellectual credit. When a researcher employs ChatGPT to draft a manuscript, the resulting text may echo ideas from sources the author has never encountered. If an AI system reproduces insights from, for example, an obscure 1975 paper without citation, does this constitute plagiarism? We argue that such cases exemplify the "provenance problem": a systematic breakdown in the chain of scholarly credit. Unlike conventional plagiarism, this phenomenon does not involve intent to deceive—researchers may disclose AI use and act in good faith—yet still benefit from the uncredited intellectual contributions of others. This dynamic creates a novel category of attributional harm that current ethical and professional frameworks fail to address. As generative AI becomes embedded across disciplines, the risk that significant ideas will circulate without recognition threatens both the reputational economy of science and the demands of epistemic justice. This Perspective analyzes how AI challenges established norms of authorship, introduces conceptual tools for understanding the provenance problem, and proposes strategies to preserve integrity and fairness in scholarly communication.

## Introduction

Suppose we ask a large language model (LLM) such as ChatGPT to help draft a section of a scholarly essay. We input a few initial prompts — fragments of an argument we're trying to shape — and the model responds with a coherent, well-articulated paragraph.

"Right!" we think to ourselves, "*That's* what we were trying to say."

There is of course a risk of self-deception here. On the one hand, perhaps the sequence of ideas in our shiny-new paragraph is, in fact, a tidied-up version of precisely what we wanted to say all along. Perhaps the argument—in this form, in this style—was there in our heads from the get-go, just half-assembled. After all, it is *we* who designed the prompts that produced this material. The model didn't conjure anything up on its own.

On the other hand, it's all a bit hazy now. If we *hadn't* used the LLM, and had instead kept on struggling with our nascent argument or phrasing, perhaps we would have come up with something substantively different. Who can say? Now that the words are on the page, we can't unsee them. They sure *look* like something we might have been trying to say…

Suppose we now decide to incorporate the paragraph into our draft (with some light edits). However, unbeknownst to us, the particular sequence and flow of ideas in this paragraph closely track those of a now-obscure paper by Smith (1975), a text we have never read. And although the LLM must have at some point 'read' it—or been trained on it, among countless other works—it might not be able to reverse-engineer its own paragraph or otherwise accurately trace it back to its source.

In any case, the LLM doesn't mention Smith (1975). Moreover, since we did not intentionally draw on Smith's work—we've never even heard of it—we don't cite it either.

Still, our essay now reflects, and perhaps even relies upon, a set of distinctive ideas that *Smith* developed and articulated half a century ago — and for which she receives no credit.

This scenario is not far-fetched. It is increasingly common for researchers to use generative artificial intelligence (AI) tools like ChatGPT to brainstorm, rephrase, summarize, or even draft material for academic work. These models are trained on vast swathes of text, including scholarly literature, but they do not (yet) always provide reliable citations or transparent sourcing.

As a result, the intellectual provenance of any given AI-generated output is often unclear—sometimes *irretrievably* so (see Box 1). This, then, raises a thorny ethical question: If we never read Smith (1975), and if the model does not, or cannot, reveal that it has drawn on her work in some opaque way, are we guilty of something akin to plagiarism? And if not, has Smith nonetheless been wronged by our failure to acknowledge her prior contribution?

---

> To properly assess the ethics of this situation, we must ask *how* the model was influenced by Smith (1975), as the mechanism creates a gradient of responsibility.
>
> **Direct input:** At one end of the spectrum, we as users might bear the most blame. Perhaps we fed the model a paper that explicitly cited Smith, or even provided Smith's article itself, and in our haste failed to properly review the sources we ourselves supplied. In such a case, the AI is tangential feature of our own clearly culpable scholarly negligence.
>
> **Retrieval systems:** Slightly more complex is the case of retrieval, where the model might use a live search tool to browse the web or a retrieval-augmented generation (RAG) architecture (Lewis et al. 2021) to query a specific database. Here, a path to attribution exists, at least in theory, and the failure to cite might be seen as a correctable flaw in the tool's design or our own due diligence.
>
> **Parametric memory:** The most profound and ethically novel problem arises from the model's parametric memory. Here, Smith's ideas surface not from any immediate source, but from the vast, latent knowledge baked into the model during its training. In this scenario, the intellectual lineage is broken in a way that may be invisible to all parties.

**Box 1: Degrees of Influence by Source and/or AI-mediation on Final Material**

At first glance, this might seem like a victimless crime. But this isn't quite right; Smith is arguably the 'victim.' It seems, rather, like a *perpetratorless* crime. After all, we did not knowingly copy anyone's work. We acted in good faith. We can even suppose that we *disclosed* our use of the LLM at the end of the essay. (We'll come back to the issue of disclosure later on.)

But plagiarism, as traditionally understood, is not just about bad faith or deliberate theft; it's about failing to credit someone for their words or distinctive ideas, despite these influencing one's own scholarship. If we benefit professionally from insights that originate with Smith, even if her influence has been mediated through a machine, and Smith receives no recognition, the underlying injustice — a misattribution of intellectual credit — arguably remains.

In this Commentary, we explore whether the use of generative AI tools in academic writing can give rise to a new class of attributional harms that don't fit neatly into existing categories of plagiarism or research misconduct, but which nonetheless challenge the moral architecture of scholarly practice.

In particular, we ask whether current norms of citation and authorship are adequate in an era when human thinkers are increasingly downstream of models trained on the intellectual labor of others, often invisibly so.

**What is plagiarism, and why is it bad?**

Plagiarism is often framed as a kind of theft — the unauthorized use of another's words or ideas (see Koplin, 2023). But in the context of academic scholarship, its moral weight stems less from property violation than from a breakdown in epistemic justice (cf. Fricker, 2007). According to this perspective, a person is wronged in her capacity as a knower if her ideas are used without proper attribution.

Of course, such a wrong may be compounded if the failure of attribution is due to bias or prejudice against the individual or her group—as in classic cases of epistemic injustice (Fricker, 2007). But even without the stain of specific prejudice, an epistemic wrong has still occurred. The original author's ideas receive no uptake as her ideas. She gets erased by omission. (Whether there is relevant bias or prejudice in our opening scenario depends on whether the LLM is an equal-opportunity source concealer, which it may or may not be.)

Put another way, on standard approaches, academic authorship functions not only to communicate knowledge but also to allocate credit for intellectual labor. This credit is reputational; it affects hiring, promotion, funding, and the broader standing of scholars in their fields. To plagiarize is, in effect, to distort the attribution of insight — to mislead readers about where ideas came from, and who deserves recognition for them.

Plagiarism's failures of attribution can also derail academic debates (Taylor 2024). Imagine that Smith's 1975 paper received some discussion in the following years. A later paper – García (1978) – identifies a subtle error in Smith (1975). Müller (1981) then refines Smith's ideas to address this criticism. By obscuring the origin of ideas, plagiarism also removes them from the surrounding critical context. Without an acknowledgement of Smith (1975) contemporary scholars would need to re-invent García's objections and Müller's refinements. At best this wastes time; at worst it leaves future work vulnerable to repeating old mistakes.

Most scholarly codes of conduct treat plagiarism as a form of research misconduct, regardless of whether the copying was intentional. The standard guidance is that authors must provide appropriate attribution when using the words, ideas, or distinctive frameworks of others. What counts as "appropriate" depends on disciplinary norms, but the underlying principle is consistent: it is unfair to represent another's intellectual labor as one's own.

The paradigmatic case of plagiarism involves knowingly copying material — perhaps from a published article, a lecture, or even an unacknowledged student paper — and presenting it without citation. But not all plagiarism is willful. Scholars can commit what is sometimes called "cryptomnesia," where they reproduce ideas they have encountered previously but mistakenly believe to be original (Taylor 1965).

Though unintentional, cryptomnesia can give rise to two distinct wrongs. The first is that someone receives academic credit that they have not earned (and with it, undeserved benefits such as jobs, promotions, or academic standing.) The second is that someone else is denied credit that they *do* deserve, and correspondingly loses out on these benefits.

But what happens when the mechanism of reproduction is not memory, but a machine?

Generative AI introduces a new wrinkle into the moral landscape of attribution. Unlike cryptomnesia, where the source of an idea has been genuinely forgotten by the human author, in the AI case the source may *never have been known* by the author.

When a scholar uses a tool like ChatGPT to help draft an essay, the model generates material derived in whole or in part from prior human work. But that derivation is often opaque. The user sees only the surface text. What lies beneath — the origin of the phrasing, the structure of the argument, the intellectual genealogy of the ideas — remains hidden.

**Failures of attribution**

This creates a novel kind of attribution problem. Traditional norms of scholarship assume a direct line of intellectual descent: we read Smith (1975), we find the ideas useful, we incorporate them (with credit) into our own argument. But in the AI-mediated case, that line is broken. The user may genuinely believe the model's output is original — or, at least, not recognizably tied to any single prior work. And yet, if the material is substantially similar to Smith's, and Smith's material is also causally implicated in the creation of the new material, then, from the perspective of epistemic credit, Smith has still been bypassed.

To be clear, the user has not committed plagiarism in the conventional sense. There is no copying with intent to deceive, or even recklessness or negligence in failing to properly "take note" of one's sources. Here, there is no *opportunity* to provide attribution to a source the user does not know exists.

But that does not mean no wrong has occurred. What matters, on this view, is that *someone else's* intellectual labor — their insight, framing, or conceptual innovation — is playing a causal role in shaping the user's work, without acknowledgment.

Of course, not all cases will be so morally charged. Much of what generative models produce is anodyne, generic, or sufficiently abstracted from any identifiable source. The boundary between individual innovation and shared intellectual substrate is often blurry.

But it is precisely this blurriness — combined with the scale and sophistication of AI-generated text — that makes the ethical stakes urgent. As scholars increasingly use these tools to accelerate their writing, synthesize literatures, or prototype arguments, we risk creating a layer of "plausible originality" that floats free of the actual minds who did the thinking.

The nature of that blurriness is itself complex, complicated by the randomness inherent in AI generation and the prominence of a source in the model's training data. The influence of a ubiquitous text, like the Bible, is strong; its frequent appearance in the training data makes verbatim reproduction likely. By contrast, a single obscure article like Smith (1975) has a far more tenuous causal link to any specific output.

It is plausible that a model, having learned a general representation of the world, could generate an idea that *coincidentally* aligns with Smith's work simply by chance—an output that might have arisen even if Smith had never written her paper. This significantly complicates the question of whether Smith was wronged or harmed.

Take a further example. A well-known paper in a niche field—say, Jones (2005)—presents a middle ground. Here, the work is prominent enough that it likely made a real causal contribution to an output that reflects its core ideas. But it is not so prominent as to be outputted verbatim, or for the role of randomness to be entirely absent (as would be the case with the Bible). This suggests a *spectrum* of attributional harm: the wrong done to Jones, whose work was probably instrumental, feels more direct than the potential wrong done to Smith, whose case may resemble a kind of algorithmic independent convergence. This element of probabilistic luck—from near-certain influence to sheer coincidence—is a defining feature of the provenance problem.

The question, then, is how to assign responsibility — and moral attention — in a context where intellectual lineage is disrupted not by bad memory or bad faith, but by a system architecture that obscures its sources to varying degrees.

**Transparency to the rescue?**

If we return to the opening scenario — where we unknowingly include in our paper a paragraph generated by ChatGPT that closely mirrors the core insights of Smith (1975) — we might wonder if we could have done something different, something to inoculate ourselves from potential wrongdoing (however inadvertent).

Maybe the trick is to be *transparent*—to *disclose* the use of generative AI in crafting the relevant section. Maybe we can go "all in" with this and exhaustively append the exact prompts we used to generate the material; the unedited text as produced by ChatGPT; and a "tracked changes" document showing the tweaks we made to that initial machine-generated material (cf. Hosseini et al. 2023).

This *would*, we think, address the issue of taking credit for an idea that one did not come up with on one's own. It would show what we specifically contributed, starting from the half-formed idea for an argument (as manifested in our initial prompts), to the final cosmetic edits we made to ChatGPT's Smith-inspired elaboration of our proto-argument.

But this solution is more complicated than it first appears. The real intellectual labor in using a generative model often lies not in the final edits, but in the iterative process of crafting the prompts themselves. One might write several paragraphs of descriptive instructions to refine a single section, meaning the work is less about editing the AI's text and more about skillfully guiding its creation. Simply appending the final prompts would fail to capture this intensive 'conversational' labor.

Conceivably, this could be fixed by uploading the *entire set* of conversations one has had with LLMs in preparing a manuscript. But doing so introduces its own serious problems. First, it creates an observer effect; scholars may work less creatively knowing their entire process is being recorded for scrutiny, much like trying to write while someone is actively watching your screen.

Second, it raises inadvertent privacy risks, especially with models that have long-term memory.

Finally, this demand for transparency quickly becomes logistical overkill. A single day's work on a paper can generate hundreds of iterative prompts. Appending such a voluminous history would be profoundly impractical for authors, reviewers, and readers alike, causing the 'solution' to collapse under its own weight.

Moreover, let's assume the final argument, as it appears on the page, is *substantially* shaped by Smith (1975). Let's assume that, without the "help" of ChatGPT, we would in fact have phrased things rather differently, or even turned out a different argument altogether. In that case, even full-on disclosure (with receipts!) would not address the issue of *Smith* failing to get credit for *her* ideas.

Transparency addresses the first wrong associated with plagiarism: our receiving unfair credit. It does not address the second: the plagiarised party failing to receive due credit.

What distinguishes this from, say, accidental convergence — where two thinkers independently arrive at similar conclusions — is the mediating role of the AI. Smith's paper may be one among millions in ChatGPT's training corpus, but it could nonetheless be doing significant unseen causal work. The fact that this work is algorithmically reprocessed rather than consciously plagiarized does not erase the loss of recognition. It simply obscures it.

To be precise, the mediating role itself is not entirely *sui generis*. One can imagine a human analogue: we discuss an idea with a senior colleague who read Smith (1975) decades ago but has long since forgotten the source. Their subconscious memory of Smith's work shapes their advice, which in turn sparks a similar insight in us. Here too, an intellectual debt is incurred through a human intermediary.

We might call this kind of scenario 'human-mediated cryptomnesia' (as distinct from AI-mediated cryptomnesia.) This older form of cryptomnesia might not be uncommon; it is a risk whenever we taken on board suggestions from colleagues, conference attendees, or peer reviewers. As in the AI case discussed above, being transparent about how we have been influenced by our colleagues, conference audiences, or peer reviewers does not address the problem of Smith failing to receive the credit she is due.

Does AI-mediated cryptomnesia raise any new issues beyond these existing human-mediated cases? One obvious difference is the potential scale of the problem. What was once an idiosyncratic risk of interpersonal exchange is now a systemic feature of an important part of our knowledge infrastructure.

The increasing ubiquity of AI writing assistance may increase the likelihood that Smith *will* be wronged. If such cases become widespread, the cumulative effect could be corrosive. Scholars whose work is absorbed into training datasets may increasingly shape the discourse without ever being cited. Those who rely heavily on AI may enjoy reputational gains built partly on others' ideas, with little incentive (or ability) to trace their provenance.

There is another difference. In cases of *human*-mediated cryptomnesia, if we are blamed for depriving Smith of credit we can reasonably try to re-allocate some of this blame to the human intermediary. If our colleague mistakenly presented Smith's ideas as his own, then perhaps they are the one who has acted wrongly (though we may still bear some responsibility for taking their contribution at face value.)

By contrast, in the case of *AI*-mediated cryptomnesia, it would make little sense to place moral blame on the AI itself. A large language model is not a moral agent; it cannot be held responsible in the way a colleague can. (Think about how absurd it would be to 'punish' a large language model for its transgressions, or of how meaningless an AI 'apology' to us would be.)

Cases of AI-mediated cryptomnesia therefore fall within what Matthias (2004) calls a 'responsibility gap' — situations where harm occurs but there is no obvious agent to whom moral blame can attach.

This evaporation of moral responsibility does not eliminate the harm to Smith. She has still been unfairly denied credit— not by us, and not (as in the human-mediated cryptomnesia) by other moral agents, but rather by a socio-technical system that divorces intellectual output from intellectual credit.

Maybe we *are* partly responsible, knowing this is a risk, for deciding to use an LLM in the first place. Nonetheless, some share of the blame – that which would normally attach itself to our human intermediary – seems to evaporate.

The risk of AI-mediated cryptomnesia don't mean that every instance of AI-assisted writing amounts to moral harm. But it does suggest that our concept of plagiarism — as a binary of guilty or innocent — may be inadequate to capture the new kinds of epistemic injustice emerging in the age of generative models.

**Doing due diligence**

If generative AI tools risk obscuring the intellectual lineage of ideas, what obligations do scholars have when using them? Traditional norms of academic writing presume a kind of traceability: when we borrow, we cite; when we adapt, we acknowledge. But these norms rely on the scholar's ability to identify the relevant source — and on the assumption that they have consciously engaged with the material. When a language model produces a passage whose intellectual ancestry is invisible, the usual mechanisms of attribution break down.

What, then, counts as due diligence?

One possibility is to place greater responsibility on the user. If scholars choose to use AI-generated text in their work, they might be expected to subject it to extra scrutiny — not just for factual accuracy, but for latent conceptual derivation.

This could involve running the text through plagiarism detection software, or treating AI outputs as drafts that require verification through conventional literature review. Scholars might leverage academic search tools like Scite, Elicit, or even LLMs themselves (Khalil & Er, 2023) to check whether AI-generated passages bear substantial similarity to existing scholarship—a kind of "originality verification" that could surface some hidden intellectual debts.

Another possibility, touched on previously, is to shift the burden to disclosure rather than detection. Scholars who use generative AI tools could be encouraged (or required) to include a statement clarifying how and to what extent such tools were used (cf. Porsdam Mann et al. 2024, Resnik & Hosseini, 2025). For example:

> "Portions of this manuscript were drafted with the assistance of ChatGPT, a generative language model. While the author did not knowingly reproduce any specific prior work, readers should be aware that the model's training data may have included material that is substantially similar to, or derivative of, published scholarship."

Such statements do not, as we have seen, solve the problem of missing attribution, but they do promote epistemic transparency. They allow readers (and peer reviewers) to understand the conditions under which the text was produced, and to interpret its originality accordingly.

One might also ask whether AI developers have obligations to improve attribution, given the moral harms associated with AI-mediated cryptomnesia. While technically challenging, such efforts could have considerable benefits for ensuring proper credit for ideas is not lost as they are absorbed into, and re-emerge from, the outputs of large language models.

**The future of writing**

The rise of generative AI tools like ChatGPT challenges our inherited assumptions about authorship, attribution, and intellectual responsibility. When scholars use these tools to produce text that may unknowingly reflect the work of others — absorbed into the model's training data and reproduced without citation — we confront a new kind of epistemic opacity. Question of plagiarism may then over-simplify things (Koplin, 2023). A scholar may act in good faith, and still benefit from the uncredited insights of others; a prior author may receive no recognition, and still have meaningfully shaped the discourse.

Of course, running underneath this analysis is the idea that human authorship and credit attribution for intellectual labour or originality should (or necessarily will) continue to drive the production of new work, career trajectories, and reputation in academia. But perhaps a different model is needed altogether (Koplin, 2023). Perhaps we should be moving toward a view of scholarship that is more collaborative and diffuse by default; that involves dynamic combinations and complex assemblages of humans and machines, where what's relevant and rewarded is—not individual credit—but the co-production of new knowledge, useful ideas, and social benefit.